\documentclass[journal]{IEEEtran}
\usepackage{verbatim}
\usepackage[centertags]{amsmath}
\usepackage{latexsym}
\usepackage{indentfirst}
\usepackage{mathrsfs}
\usepackage{amsthm}
\usepackage{amssymb}
\usepackage{amsfonts}
\usepackage{amsbsy}
\usepackage{url}
\usepackage{tikz}
\usepackage{multirow}
\usepackage{array}
\usepackage{pdflscape}
\usepackage{float}
\usepackage{booktabs}

\newtheoremstyle{mytheorem}%                % Name
  {}%                                     % Space above
  {}%                                     % Space below
  {\itshape}%                                     % Body font
  {}%                                     % Indent amount
  {}%                                       % Theorem head font
  {.}%                                    % Punctuation after theorem head
  { }%                                    % Space after theorem head, ' ', or \newline
  {}%                                     % Theorem head spec (can be left empty, meaning `normal')

\newtheorem{theorem}{Theorem}

\newtheorem{proposition}[theorem]{Proposition} 

\theoremstyle{definition}

\newtheorem{example}[theorem]{Example}

\newcommand{\Fq}{\mathbb{F}_{q}}

\newcommand{\F}{{\mathbb F}}
\newcommand{\G}{{\mathbb G}}
\newcommand{\HH}{{\mathbb H}}
\newcommand{\Tr}{{\rm Tr}}

\newcommand{\Span}{{\rm Span}}

\DeclareMathOperator{\wt}{wt}

\newcommand{\etal}{\emph{et al. }}
\newcommand{\eg}{\emph{e.g.}}
\newcommand{\ie}{\emph{i.e.}}

\allowdisplaybreaks

\begin{document}

\title{Good Stabilizer Codes from Quasi-Cyclic Codes over $\F_4$ and $\F_9$}

\author{Martianus Frederic Ezerman, San Ling, Buket \"{O}zkaya, and Patrick Sol\'e
	\thanks{M.~F.~Ezerman, S.~Ling, and B.~\"{O}zkaya are with the Division of Mathematical Sciences, School of Physical and Mathematical Sciences, Nanyang Technological University, 21 Nanyang Link, Singapore 637371, e-mails: ${\rm\{fredezerman,\, lingsan,\,buketozkaya\}}$@ntu.edu.sg.}
	\thanks{P. Sol\'e is with Aix-Marseille Universit\'e, CNRS, Centrale Marseille, I2M, Marseille, France,
		e-mail: sole@enst.fr.}
	\thanks{
		M.~F.~Ezerman, S.~Ling, and B.~\"{O}zkaya are supported by Nanyang Technological University Research Grant M4080456.}
	\thanks{The authors thank Markus Grassl for his support in providing a wide database of best-known qubit and qutrit stabilizer codes.}
	\thanks{This work has been accepted for presentation at the International Symposium on Information Theory ISIT 2019. Copyright (c) 2019 IEEE. Personal use of this material is permitted. However, permission to use this material for any other purposes must be obtained from the IEEE by sending a request to pubs-permissions@ieee.org.}
}
\maketitle

\begin{abstract}
We apply quantum Construction X on quasi-cyclic codes with large Hermitian hulls over $\F_{4}$ and $\F_9$ to derive good qubit and qutrit stabilizer codes, respectively. In several occasions we obtain quantum codes with stricly improved parameters than the current record. In numerous other occasions we obtain quantum codes with best-known performance. For the qutrit ones we supply a systematic construction to fill some gaps in the literature.
\end{abstract}


\section{Introduction}\label{intro}

Turning large scale quantum computing into practical reality remains a huge challenge to engineer. Keeping the errors in the system below the fidelity treshold is key since noise, if can be kept below a certain level, is not an obstacle to resilient quantum computation~\cite{Knill1998}. The possibility of correcting errors in the qubit systems was shown, \eg, in the early works of Shor \cite{Shor1995}, Steane \cite{Steane1996} and Laflamme \etal \cite{Laflamme1996}. Developments in this active topic up to 2011 is well-documented in~\cite{Lidar2013}. Advances continue to be made as effort intensifies in the race to be among the first to make quantum computing scalable. %\new{We limit our references to the vast literature on quantum error control only to those directly relevant to our construction.}

Let $\Fq$ denote the finite field with $q$ elements, where $q$ is a prime power. 
The {\it stabilizer formalism}, discussed in Gottesman's thesis \cite{Gottesman97} and, in the qubit (quantum bit) case, described in the language of group algebra by Calderbank \etal in \cite{Calderbank1998}, remains the most widely-studied model of error control. Ketkar \etal generalized the formalism to $q$-ary quantum codes derived from classical codes over $\F_{q^2}$ in~\cite{Ketkar}. 

Let $V_{n}=(\mathbb{C}^{q})^{\otimes n}$ be the $n$-fold tensor power of $\mathbb{C}^{q}$. A \textit{$q$-ary  quantum code} of length $n$ is a subspace $\mathcal{Q}$ of $V_{n}$ with dimension $K \geq 1$. If the code $\mathcal{Q}$ is a stabilizer code, we use the notation $\mathcal{Q}=[\![n,k,d]\!]_q$, where $k=\log_q K$. The {\it propagation rules} in the next proposition will be useful later.

\begin{proposition}(\cite[Thm. 6]{Calderbank1998}) \label{propagation}Let $\mathcal{Q}$ be an $[\![n,k,d]\!]_q$ code. Then the following quantum codes exist. 
%\edit{Check if the rule based on the spectra can be easily applied here. If yes, then add it to the enumeration below.}
\begin{enumerate}
	\item[i)] $[\![n,k-1, \geq d]\!]_q$ (by subcode construction).
	\item[ii)] $[\![n+1,k, \geq d]\!]_q$ (by lengthening).
	\item[iii)] $[\![n-1,k, \geq d-1]\!]_q$ (by puncturing).
\end{enumerate}
\end{proposition} 

Lisonek and Singh proposed an interesting modification to the construction of qubit stabilizer codes by relaxing the self-orthogonality requirement in~\cite{Lisonek2014}. Their framework, inspired by Construction X in the classical setup (see~\cite[Chapter 18 \S 7.1]{MS78}), yielded a number of better qubit codes than the previous best-known. These better codes came from applying their construction to specifically chosen cyclic codes over $\F_4$. 

Construction X for qubit codes generalizes naturally to $q$-ary quantum codes. The case of $p$-ary for primes $p$ was shown in~\cite[Thm.~4]{Degwekar2015}. The first statement in the proof of \cite[Prop.~1]{Lisonek2014} holds for any $q$ since the trace mapping $\mathrm{Tr}_{\F_{q^2}/\F_q}$ is onto. The construction of the orthonormal set $B$ in \cite[Prop.~1]{Lisonek2014} can then be replicated and the general version of~\cite[Thm.~2]{Lisonek2014} follows from the exact same argument as in the original version.

It is unfortunate that the $p$-ary codes, with $p>2$, in~\cite[Table 1]{Degwekar2015} are compared to qubit codes. The authors' claim that the obtained $p$-ary quantum codes improve on the parameters of known qubit codes is incorrect. The codes being compared live in different universes. They represent, respectively, quantum error operators on different Hilbert spaces, namely $(\mathbb{C}^{p})^{\otimes n}$ and $(\mathbb{C}^{2})^{\otimes n}$, and, thus, are incomparable to each other.

This work uses self-orthogonal or nearly self-orthogonal quasi-cyclic codes over $\F_4$ and $\F_9$ as ingredients in the quantum Construction X to derive good qubit and qutrit ($3$-ary quantum) codes. Such codes have a good chance to be implemented in actual quantum processors. We exhibit quantum codes with parameters that \emph{strictly} improve on the currently best-known ones and list those that match the best-known ones in performance.

Let $\vec{v}:=(v_1,\ldots,v_n)$ and $\vec{u}:=(u_1,\ldots,u_n)$ be vectors in $\F_{q^2}^n$. Their {\it Hermitian inner product} $\left\langle \vec{v}, \vec{u} \right\rangle_{\rm H}$ is $\sum_{i=1}^n v_i u_i^q$. The {\it weight} of $\vec{v}$, denoted by $\wt(\vec{v})$, is the number of nonzero entries in $\vec{v}$. A linear code $C$ over $\F_{q^2}$ of length $n$, dimension $k$ and minimum distance $d:=d(C)$ is denoted by $[n,k,d]_{q^2}$. We let $A+B:=\{\vec{v} + \vec{u} : \vec{v} \in A, \vec{u} \in B\}$, for subspaces $A$ and $B$ in $\F_{q^2}^n$, and use $\dim(A)$ to abbreviate $\dim_{\F_{q^2}}(A)$.

Instead of starting with an $[n,k,d]_{q^2}$-code $C$ that contains its Hermitian dual $C^{\bot_{\rm H}}$ with parameters $[n,n-k,d^{\bot}]_{q^2}$, we will begin with a Hermitian self-orthogonal $[n,k,d]_{q^2}$-code $C$, \ie, $C \subseteq C^{\bot_{\rm H}}$. We end this introduction by restating quantum Construction X in the form that matches our preference.

\begin{theorem}(\cite[Thm. 2]{Lisonek2014} and \cite[Thm. 4]{Degwekar2015})\label{thm:X}. 
For an $[n,k]_{q^2}$-linear code $C$, let $e:=k - \dim(C \cap C^{\bot_{\rm H}})$. Then there exists an $[\![n+e,n-2k+e,d(\mathcal{Q})]\!]_q$ quantum stabilizer code $\mathcal{Q}$ with $d(\mathcal{Q}) \geq \min \{d(C^{\bot_{\rm H}}), d(C + C^{\bot_{\rm H}})+1\}$. 
\end{theorem} 

Note that $e=0$ describes the usual stabilizer construction. Stabilizer codes from QC codes of index $2$ was discussed in~\cite{Galindo2018}. To avoid a sharp drop in $d$, we would like $e$ to be small, \ie, the {\it Hermitian hull} $C \cap C^{\bot_{\rm H}}$ to be large. The search reported in \cite{Lisonek2014} on the family of linear cyclic codes over $\F_{4}$ found improved qubit codes with $e \in \{1,2,3\}$.

\section{Quasi-Cyclic Codes}\label{QC}
Let $m$ and $\ell$ be positive integers such that $\gcd(m,q)=1$. A linear code $C$ of length $m\ell$ over $\Fq$ is called a quasi-cyclic (QC) code of index $\ell$ if it is invariant under shift of codewords by $\ell$ positions and $\ell$ is the minimal number with this property. The code $C$ is cyclic when $\ell=1$. If we view any codeword of $C$ as an $m \times \ell$ arrays 
\begin{equation}\label{array}
\vec{c}=\left(
  \begin{array}{ccc}
    c_{0,0} & \ldots & c_{0,\ell-1} \\
    \vdots & \vdots  & \vdots \\
    c_{m-1,0} & \ldots & c_{m-1,\ell-1} \\
  \end{array}
\right),
\end{equation} 
then being invariant under shift by $\ell$ units in $\Fq^{m\ell}$ amounts to being closed under row shift in $\Fq^{m\times\ell}$.

Consider the principal ideal $I=\langle x^m-1 \rangle$ of $\Fq[x]$ and define the quotient ring $R:=\Fq[x]/I$. We associate a vector $\vec{c} \in \Fq^{m\times \ell} \simeq \Fq^{m\ell}$ as in (\ref{array}) with an element of $R^\ell$ as
\begin{equation} \label{associate-1}
\vec{c}(x):=(c_0(x),c_1(x),\ldots ,c_{\ell-1}(x)) \in R^\ell ,
\end{equation}
where for each $0\leq t \leq \ell-1$, 
\begin{equation}\label{columns} 
c_t(x):= c_{0,t}+c_{1,t}x+c_{2,t}x^2+\ldots +
c_{m-1,t}x^{m-1} \in R.
\end{equation} 
Then, the following map is an $R$-module isomorphism.
\begin{equation}\label{identification-1}
\begin{array}{ccc} 
\phi: \F_q^{m\ell} & \longrightarrow & R^\ell  \\
\vec{c} & \longmapsto & \vec{c}(x) .
\end{array}
\end{equation}
 For $\ell=1$, this is the classical polynomial representation of cyclic codes. Note that the row shift in $\Fq^{m\times\ell}$ corresponds to componentwise multiplication by $x$ in $R^\ell$. Thus, a $q$-ary QC code $C$ of length $m\ell$ and index $\ell$ is an $R$-submodule in $R^{\ell}$.

%We now describe the decomposition of 
A QC code over $\F_q$ decomposes into shorter codes over field extensions of $\F_q$. Further details are given in~\cite{LS01}. The self-reciprocal polynomial $x^m-1$ factors into irreducible polynomials in $\F_q[x]$ as
\begin{equation}\label{irreducibles}
x^{m}-1=g_{1}(x)\cdots g_{s}(x) \, h_{1}(x) \, h_{1}^*(x)\cdots h_{r}(x) \,h_{r}^*(x),
\end{equation}
where $g_{i}$'s are self-reciprocal and $h_{j}^*$ denotes the reciprocal of $h_{j}$, for all $i,j$. Since $\gcd(m,q)=1$, there are no repeating factors in (\ref{irreducibles}). By the Chinese Remainder Theorem (CRT), 
\begin{equation} \label{CRT-1}
\small{R \cong \bigoplus_{i=1}^{s} \F_q[x]/\langle g_{i}(x) \rangle  \oplus
\bigoplus_{j=1}^{r} \Bigl( \F_q[x]/\langle h_{j}(x) \rangle \oplus \F_q[x]/\langle h_{j}^*(x) \rangle \Bigr).}
\end{equation}\normalsize
Since each $g_i(x)$, $h_{j}(x)$ and $h_{j}^*(x)$ divides $x^m-1$, their roots are powers of some fixed primitive $m^{\text{th}}$ root of unity $\xi$. For each $i\in \{1,\ldots ,s\}$, let $u_i$ be the smallest nonnegative integer such that $g_i(\xi^{u_i})=0$. For each $j\in \{1,\ldots, r\}$, let $v_j$ be the smallest nonnegative integer such that $h_j(\xi^{v_j})=h_{j}^*(\xi^{-v_j})=0$. Since all factors in (\ref{irreducibles}) are irreducible, direct summands in (\ref{CRT-1}) are isomorphic to field extensions of $\F_q$. Let $\G_{i}:=\F_q(\xi^{u_i})\simeq\F_q[x]/\langle g_{i}(x) \rangle$, $\HH_{j}':=\F_q(\xi^{v_j})\simeq\F_q[x]/\langle h_{j}(x) \rangle$ and $\HH_{j}'':=\F_q(\xi^{-v_j})\simeq\F_q[x]/\langle h_{j}^*(x) \rangle$ denote those extensions, for each $i$ and $j$, respectively. By CRT, the decomposition of $R$ in (\ref{CRT-1}) now becomes
\begin{align}\label{CRT-2}
R &\cong \bigg( \bigoplus_{i=1}^{s} \G_{i} \bigg) \oplus \bigg( \bigoplus_{j=1}^{r} \Big( \HH_{j}' \oplus \HH_{j}'' \Big) \bigg)\\
f(x) &\mapsto \left(\big[ f(\xi^{u_i})\big]_{1\leq i\leq s}, \big[ f(\xi^{v_j})\big]_{1\leq j\leq r}, \big[ f(\xi^{-v_j})\big]_{1\leq j\leq r}\right) \notag.
\end{align}
This implies that
\begin{align} \label{CRT-3}
R^{\ell} & \cong  \bigg(\bigoplus_{i=1}^{s} \G_i^{\ell}\bigg) \oplus \bigg(\bigoplus_{j=1}^{r} (\HH'_{j})^{\ell} \oplus (\HH''_{j})^{\ell}\bigg)\\
\vec{f}(x) &\mapsto \left(\big[ \vec{f}(\xi^{u_i})\big]_{1\leq i\leq s}, \big[ \vec{f}(\xi^{v_j})\big]_{1\leq j\leq r}, \big[ \vec{f}(\xi^{-v_j})\big]_{1\leq j\leq r}\right), \notag
\end{align}
where $\vec{f}(a)$ denotes the componentwise evaluation at $a$, for any element $\vec{f}(x)=\bigl(f_{0}(x),\ldots ,f_{\ell-1}(x)\bigr)\in R^{\ell}$. Hence, as an $R$-submodule of $R^{\ell}$, a QC code $C\subseteq R^{\ell}$ decomposes into 
\begin{equation} \label{constituents}
C\cong\bigg( \bigoplus_{i=1}^{s} C_i \bigg) \oplus \bigg( \bigoplus_{j=1}^{r} \Bigl( C_{j}' \oplus C_{j}'' \Bigr) \bigg).
\end{equation}
Here $C_i$'s are the $\G_i$-linear codes of $C$ of length $\ell$, for all $i=1,\ldots,r$, while $C'_{j}$'s and $C''_{j}$'s are the $\HH'_j$- and $\HH''_j$-linear codes of $C$ of length $\ell$, respectively, for all $j=1,\ldots,r$. We call these linear codes of length $\ell$ over various extensions of $\F_q$ the {\it constituents} of $C$.

Let $C\subseteq R^\ell$ be generated by $\{\vec{f}_1(x),\ldots, \vec{f}_n(x)\}$, where $\vec{f}_b(x)=\bigl(f_{b,0}(x),\ldots ,f_{b,\ell-1}(x)\bigr)\in R^{\ell}$, for each $1\leq b \leq n$. Then, for $1\leq i \leq s$ and $1\leq j \leq r$, we have
\begin{align}\label{explicit constituents}
C_i &= \Span_{\G_i}\bigl\{\vec{f}_b(\xi^{u_i}) : 1\leq b \leq n\bigr\},  \notag\\
C'_{j} &=\Span_{\HH'_j}\bigl\{\vec{f}_b(\xi^{v_j}) : 1\leq b \leq n \bigr\},  \\  \notag
C''_{j} &= \Span_{\HH''_j}\bigl\{\vec{f}_b(\xi^{-v_j}) : 1\leq b \leq n \bigr\}.  
\end{align}

Conversely, let $C_i\subseteq\G_i^{\ell}$, $C'_j\subseteq\HH_j'^{\ell}$ and $C''_j\subseteq\HH_j''^{\ell}$ be arbitrary linear codes, for each $i\in \{1,\ldots ,s\}$ and each $j\in \{1,\ldots, r\}$, respectively. Then, by \cite[Thm.~5.1]{LS01}, an arbitrary codeword $\vec{c}$ in the corresponding $q$-ary QC code $C$ described as in (\ref{constituents}) can be written as an $m\times \ell$ array like in (\ref{array}) such that each row of $\vec{c}$ is of the form
\begin{equation}\label{trace codeword}
\begin{aligned}
\vec{c}_g & = \frac{1}{m} \bigg(\sum\limits_{i=1}^{s} \Tr_{\G_{i}/\F_q} \big( \lambda_{i,t} \xi^{-g u_{i}} \big) + \bigg. \\
&\bigg. \sum\limits_{j=1}^{r} \left[\Tr_{\HH_j'/\F_q} \big(\lambda'_{j,t}\xi^{-gv_j} \big) +\Tr_{\HH_j''/\F_q} \big(\lambda''_{j,t}\xi^{gv_j} \big)\right] \bigg),
\end{aligned}
\end{equation}
for $0\leq g \leq m-1$, where $\vec{\lambda}_i=(\lambda_{i,0},\ldots ,\lambda_{i,\ell-1}) \in C_i$, for all $i$, $\vec{\lambda}_j'=(\lambda'_{j,0},\ldots ,\lambda'_{j,\ell-1}) \in C_j'$ and $\vec{\lambda}_j''=(\lambda''_{j,0},\ldots ,\lambda''_{j,\ell-1}) \in C_j''$, for all $j$. Since $\frac{1}{m}C=C$, we can cancel $\frac{1}{m}$ out. Note that, in this representation, the row shift invariance of codewords amounts to being closed under multiplication by $\xi^{-1}$.

The cardinality, say $q_i$, of each $\G_i$ is an even power of $q$. Each $\G_i^{\ell}$ is equipped with the Hermitian inner product. For $1\leq j \leq r$, $\HH_j'^{\ell}$ and $\HH_j''^{\ell}$ are equipped with the usual Euclidean inner product. Observe that $\HH_j'=\HH_j''$ follows from the fact that $\Fq(\alpha\xi^a)=\Fq(\alpha\xi^{-a})$, for any $a\in\{0,\ldots,m-1\}$. 

The dual of a QC code is also QC. For the proof of the following result we refer to \cite{LS01}.
\begin{proposition}\label{duality}
Let $C$ be a QC code with CRT decomposition as in (\ref{constituents}). Then 
%its (Euclidean or Hermitian) dual code 
$C^{\bot_{\rm H}}$ is of the form
\begin{equation}\label{dual}
C^{\bot_{\rm H}}=\bigg( \bigoplus_{i=1}^s C_i^{\bot_{\rm H}} \bigg) \oplus 
\bigg( \bigoplus_{j=1}^r \Bigl( C_j''^{\bot_{\rm E}} \oplus C_j'^{\bot_{\rm E}} \Bigr)\bigg),
\end{equation}
where  $\bot_{\rm H}$ denotes the Hermitian dual on $G_i^\ell$ 
%(for all $1\leq i \leq s$) 
and  $\bot_{\rm E}$ denotes the Euclidean dual on $\HH_j'^{\ell}=\HH_j''^{\ell}$.
% (for all $1\leq j \leq r$).
\end{proposition}

By (\ref{constituents}) and (\ref{dual}), we characterize self-orthogonal QC codes.
\begin{theorem} \label{SO criteria}
Let $C$ be a $q$-ary QC code of length $m\ell$ whose CRT decomposition is as in (\ref{constituents}). Then $C$ is Hermitian self-orthogonal 
%(with respect to the Euclidean or Hermitian inner product) 
if and only if $C_i$ is Hermitian self-orthogonal over $\G_i$, for all $1\leq i \leq s$, and $C_t'' \subseteq C_j'^{\bot_{\rm E}}$ (or equivalently $C_t' \subseteq C_j''^{\bot_{\rm E}}$) over $\HH_j'=\HH_j''$, for all $1\leq j \leq r$.
\end{theorem}

\section{Designing the Hermitian Hull}\label{hull section}

From this point on we consider Hermitian self-orthogonal or nearly self-orthogonal QC codes over $\F_{q^2}$. As was done in Theorem \ref{SO criteria} for self-orthogonality, the CRT decompositions of the QC code $C$ in (\ref{constituents}) and of its Hermitian dual $C^{\bot_{\rm H}}$ in (\ref{dual}) characterize nearly self-orthogonal QC codes. 

Let $k$ denote the dimension of $C$ over $\F_{q^2}$. Clearly, $C^{\bot_{\rm H}}$ has dimension $m\ell-k$. Let $e_i:=[\G_i : \F_{q^2}]=\deg (g_i(x))$ and $k_i:=\dim_{\G_i} (C_i)$, for all $1\leq i \leq s$. Let $e_j:=[\HH_j' : \F_{q^2}]=[\HH_j'' : \F_{q^2}]=\deg( h_j'(x))=\deg (h_j''(x))$, $k_j':=\dim_{\HH_j'} (C_j')$ and $k_j'':=\dim_{\HH_j''} (C_j'')$, for all $1\leq j \leq r$. Then we have 
\begin{equation}\label{dimension}
m =\sum_{i=1}^s e_i + \sum_{j=1}^r 2 e_j,\ k =  \sum_{i=1}^s e_i k_i + \sum_{j=1}^r  e_j  (k_j' + k_j'').
\end{equation}
Clearly, $x-1$ is one of the self-reciprocal divisors of $x^m-1$, for any positive integer $m$, and $x+1$ is another such divisor if $m$ is even. Recall that $x-1$ and $x+1$ coincide whenever $q$ is even. WLOG, let $g_1(x):=x\pm 1$. We have $\G_1=\F_{q^2}$ and rewrite $k$ in (\ref{dimension}) as \vspace{-5pt}
\begin{equation}\label{X-dimension}
k = \dim (C_1) + \sum\limits_{i=2}^s e_i k_i + \sum_{j=1}^r  e_j   (k_j' + k_j'').
\end{equation}
By using (\ref{X-dimension}) and Proposition \ref{duality}, we obtain \vspace{-5pt}
\begin{multline}\label{hull-dimension}
\hspace{-10pt}\dim (C\cap C^{\bot_{\rm H}})  = \dim (C_1 \cap C_1^{\bot_{\rm H}}) + \sum_{i=2}^s e_i \dim_{\G_i} (C_i \cap C_i^{\bot_{\rm H}}) + \\ 
 \sum_{j=1}^t e_j \big( \dim_{\HH_j'} (C_j' \cap C_j''^{\bot_{\rm E}}) + \dim_{\HH_j''} (C_j'' \cap C_j'^{\bot_{\rm E}})\big) .
\end{multline}
Assuming $C$ to be Hermitian self-orthogonal is equivalent to saying that $C=C\cap C^{\bot_{\rm H}}$, where (\ref{X-dimension}) and (\ref{hull-dimension}) also coincide. To use Theorem \ref{thm:X} with QC codes in the desired way, we assume that all constituent codes except $C_1$ satisfy the requirements of Theorem \ref{SO criteria}. The code $C_1$ has a bit more freedom since we set $e:=\dim(C_1) - \dim(C_1 \cap C_1^{\bot_{\rm H}})$. %, where $e \in \{0,1,2\}$. 
If the remaining constituents $C_i,C_j',C_j''$ agree with the conditions in Theorem \ref{SO criteria}, for all $2\leq i \leq s$ and $1\leq j \leq r$, then we obtain $e=k - \dim(C \cap C^{\bot_{\rm H}})$ easily by subtracting (\ref{hull-dimension}) from (\ref{X-dimension}). The next section presents the outcomes of a random search over such constituent codes and the resulting stabilizer codes.

\section{Good Qubit and Qutrit Codes}

%%%% Table 1
\begin{table*}[t!] 
	\caption{Explicit constructions for qutrit stabilizers with best-known parameters}
	\renewcommand{\arraystretch}{1.2}
	\centering
	\begin{tabular}{c|l|ccc|cc || c|l|ccc|cc || c|l|ccc|cc  }
		\hline
		No & Stabilizer & $e$ & $\ell$ & $m$ & $s$ & $r$  &
		No & Stabilizer & $e$ & $\ell$ & $m$ & $s$ & $r$  &
		No & Stabilizer & $e$ & $\ell$ & $m$ & $s$ & $r$ \\
		\hline
		1 & $[\![4, 0, 3]\!]_3$ & 0 & 2 & 2 & 2 & 0 &
		16 & $[\![15, 9, 3]\!]_3$ & 0 & 3 & 5 & 3 & 0 &
		31 & $[\![33, 23, 4]\!]_3$ & 1 & 2 & 16 & 2 & 5 \\
		
		2 & $[\![8, 0, 4]\!]_3$ &  &  & 4 & 2 & 1 & 
		17 & $[\![30, 24, 3]\!]_3$ &  &  & 10 & 6 & 0 &
		32 & $[\![7, 3, 3]\!]_3$ &  & 3 & 2 & 2 & 0 \\
		
		3 & $[\![8, 2, 4]\!]_3$ &  &  & 4 & 2 & 1 &
		18 & $[\![8, 4, 3]\!]_3$ &  & 4 & 2 & 2 & 0 &
		33 & $[\![13, 5, 4]\!]_3$ &  &  & 4 & 2 & 1\\
		
		4 & $[\![10, 6, 3]\!]_3$ &  &  & 5 & 3 & 0 &
		19 & $[\![10, 2, 4]\!]_3$ &  & 5 & 2 & 2 & 0 &
		34 & $[\![13, 7, 3]\!]_3$ &  &  & 4 & 2 & 1 \\
		
		5 & $[\![16, 6, 4]\!]_3$ &  &  & 8 & 2 & 3 &
		20 & $[\![25, 15, 4]\!]_3$ &  &  & 5 & 3 & 0 &
		35 & $[\![31, 13, 6]\!]_3$ &  &  & 10 & 6 & 0 \\
		
		6 & $[\![16, 8, 4]\!]_3$ &  &  & 8 & 2 & 3 &
		21 & $[\![25, 19, 3]\!]_3$ &  &  & 5 & 3 & 0 &
		36 & $[\![31, 21, 4]\!]_3$ &  &  & 10 & 6 & 0 \\
		
		7 & $[\![20, 4, 6]\!]_3$ &  &  & 10 & 6 & 0 &
		22 & $[\![5, 1, 3]\!]_3$ & 1 & 2 & 2 & 2 & 0 &
		37 & $[\![11, 3, 4]\!]_3$ &  & 5 & 2 & 2 & 0 \\
		
		8 & $[\![20, 6, 6]\!]_3$ &  &  & 10 & 6 & 0 &
		23 & $[\![9, 3, 3]\!]_3$ &  &  & 4 & 2 & 1 &
		38 & $[\![26, 16, 4]\!]_3$ &  &  & 5 & 3 & 0 \\
		
		9 & $[\![20, 8, 5]\!]_3$ &  &  & 10 & 6 & 0 &
		24 & $[\![11, 5, 3]\!]_3$ &  &  & 5 & 3 & 0 & 
		39 & $[\![12, 4, 4]\!]_3$ & 2 & 2 & 5 & 3 & 0 \\
						
		10 & $[\![20, 10, 4]\!]_3$ &  &  & 10 & 6 & 0 &
		25 & $[\![17, 3, 6]\!]_3$ &  &  & 8 & 2 & 3 &
		40 & $[\![18, 6, 5]\!]_3$ &  &  & 8 & 2 & 3 \\
						
		11 & $[\![20, 12, 4]\!]_3$ &  &  & 10 & 6 & 0 &
		26 & $[\![21, 5, 6]\!]_3$ &  &  & 10 & 6 & 0 &
		41 & $[\![18, 8, 4]\!]_3$ &  &  & 8 & 2 & 3 \\

		12 & $[\![20, 14, 3]\!]_3$ &  &  & 10 & 6 & 0 &		
		27 & $[\![21, 9, 5]\!]_3$ &  &  & 10 & 6 & 0 &
		42 & $[\![18, 12, 3]\!]_3$ &  &  & 8 & 2 & 3 \\
		
		13 & $[\![22, 12, 4]\!]_3$ &  &  & 11 & 1 & 1 &
		28 & $[\![21, 13, 4]\!]_3$ &  &  & 10 & 6 & 0 &
		43 & $[\![22, 8, 5]\!]_3$ &  &  & 10 & 6 & 0 \\
		
		14 & $[\![28, 18, 4]\!]_3$ &  &  & 14 & 2 & 2 & 
		29 & $[\![23, 11, 5]\!]_3$ &  &  & 11 & 1 & 1 &
		44 & $[\![22, 12, 4]\!]_3$ &  &  & 10 & 6 & 0 \\

		15 & $[\![12, 6, 3]\!]_3$ &  & 3 & 4 & 2 & 1 &
		30 & $[\![33, 15, 6]\!]_3$ &  &  & 16 & 2 & 5 &		
		45 & $[\![22, 16, 3]\!]_3$ &  & 5 & 4 & 2 & 1 \\
		\hline
	\end{tabular}
	\label{tab:qutrit}
\end{table*}
%%%%%

For qubit, the most comprehensive record is Grassl's online table \cite{Grassl}. Systematically constructed (\ie, not in the form of stored values) stabilizer matrices corresponding to the currently best codes in $(\mathbb{C}^{2})^{\otimes n}$, for $n \leq 96$, are available for many entries. It is a two-fold challenge to contribute meaningfully. First, for $n \leq 100$, many researchers have attempted exhaustive search. Better codes are unlikely to be found without additional clever strategy. Second, for $n > 100$, computing the actual distance $d(\mathcal{Q})$ tends to be prohibitive.

Less attention has been given to qutrit codes, for which there is no publicly available database of comparative extense. A table listing numerous qutrit codes is kept by Y. Edel in~\cite{Edel} based on their explicit construction as {\it quantum twisted codes} in~\cite{BE}. Better codes than many of those in the table have since been found. M. Grassl generously allowed us access to his offline database that contains best-known qutrit codes up to length $50$ for comparison. In most of the nontrivial cases, the stabilizer matrices are stored matrices. Adding systematic ways to build them remains a valuable endeavour here. 

\begin{example}
For qubits, our search yields a $[\![31,9,7]\!]_2$ code, which is \emph{strictly better} than the prior best-known $[\![31,9,6]\!]_2$ code. Let us describe a quaternary QC code, which gives rise to the better code. Let $\zeta$ be a primitive element in $\F_4$, $m=15$ and $\ell=2$. Over $\F_4$, the factorization of $x^{15}-1$ (cf. (\ref{irreducibles})) is 
\begin{multline*}
\hspace{-10pt}x^{15}-1 \hspace{-2pt} =
\hspace{-2pt}(x + 1)(x^2+\zeta x+1)(x^2+\zeta^2x+1)[(x+\zeta)(x+\zeta^2)]\\ \hspace{22pt}\big[(x^2+x+\zeta)(x^2+\zeta^2x+\zeta^2)\big]\big[(x^2+x+\zeta^2)(x^2+\zeta x+\zeta)\big].
\end{multline*}
The first $3$ factors are self-reciprocal and the remaining $6$ factors are ordered in reciprocal pairs. Hence, we have $s=r=3$ such that $\G_1=\F_4$, $\G_2=\G_3=\F_{16}$, $\HH'_1=\HH''_1=\F_4$ and $\HH'_j=\HH''_j=\F_{16}$, for $j\in\{2,3\}$. Assuming this ordering, consider the following constituents of length $2$ with their generator matrices corresponding to the ordered factors: 
\begin{align}\label{qubit cons}
&C_1 : \begin{pmatrix}  0 & 1 \end{pmatrix}, \, C_2 = C_3 : \begin{pmatrix}  1 & \omega \end{pmatrix}, \,
C_4 : 0_2, \, C_5 : \begin{pmatrix}  1 & \xi^{10} \end{pmatrix},\notag\\
&C_1' :  \begin{pmatrix}  1 &  1 \end{pmatrix}, \, C_1'' : I_2, \, C_2' : 0_2, \, C_2'' : I_2,
\end{align}
where $\xi$ is a primitive element of $\F_{16}$ satisfying $\xi^2+\xi+\zeta=0$, $0_2:=\begin{pmatrix}  0 & 0 \end{pmatrix}$, and $I_2$ denotes the $2\times 2$ identity matrix.

The QC code $C\subseteq\F_4^{30}$ of index $2$ with those constituents has dimension $11$, by (\ref{dimension}). If $R=\F_4[x]/\langle x^{15}-1\rangle$, then $C$ is generated by $\vec{f}_1(x)=(f_{1,0}(x), f_{1,1}(x))$ and $\vec{f}_2(x)=(f_{2,0}(x), f_{2,1}(x))$ as an $R$-submodule in $R^2$, with 
\begin{align*}
f_{1,0}(x)=&~\zeta x^{12} + x^{10} + \zeta^2 x^9 + \zeta^2 x^6 + x^5 + \zeta x^3, \\
f_{1,1}(x)=&~\zeta^2 x^{13} + \zeta^2 x^{12} + \zeta^2 x^{10} + x^9 + \zeta^2 x^7 + x^6 +\\ &~\zeta^2 x^5 + \zeta x^4 + \zeta^2 x^3 + \zeta x + 1,\\
f_{2,0}(x)=&~0,\\
f_{2,1}(x)=&~\zeta^2 x^{14} + \zeta x^{13} + \zeta^2 x^{12} + \zeta^2 x^{11} + \zeta x^9 + x^8 +\\ &~\zeta x^7 + \zeta x^6 + x^4 + \zeta^2 x^3 + x^2 + x.
\end{align*}
The generator polynomials are found by applying (\ref{trace codeword}) on the constituents given in (\ref{qubit cons}) followed by the map $\phi$ in (\ref{identification-1}). Conversely, one obtains the constituents listed in (\ref{qubit cons}) by evaluating $\vec{f}_1(x)$ and $\vec{f}_2(x)$ at $1, \xi^6, \xi^3, \zeta, \zeta^2, \xi, \xi^{11}, \xi^2,\xi^7$, respectively. It is easy to verify that all constituents except $C_1$ satisfy the requirements of Theorem \ref{SO criteria}, whereas $\dim (C_1 \cap C_1^{\bot_{\rm H}})=0$, implying $e=1$. Hence, $C \cap C^{\bot_{\rm H}}$ has dimension $10$. The Hermitian dual of $C$ is a $[30,19,7]_4$ code, which attains the best-known distance for a quaternary code of this length and dimension, and $d(C + C^{\bot_{\rm H}})=6$. Thus, by Theorem \ref{thm:X}, we obtain a $[\![31,9,7]\!]_2$ stabilizer code.

The propagation rules in Proposition \ref{propagation} give codes with parameters $[\![31,8,7]\!]_2$, $[\![32,9,7]\!]_2$ and $[\![30,9,6]\!]_2$. Their performance matches the current best. 
\end{example}

\begin{example}
In the qutrit case we obtain an \emph{optimal} $[\![17,7,5]\!]_3$ code whose distance reaches the upper bound. The previous best-known was a $[\![17,7,4]\!]_3$ code. Let us explain how to obtain the optimal code. Let $\omega$ be a primitive element in $\F_9$, $m=8$ and $\ell=2$. The decomposition of $x^{8}-1$ into linear factors over $\F_9$ is
\[
x^8-1=\prod_{j=0}^{7} (x+\omega^j)=(x+1)(x-1)\prod_{j=1}^{3} \big[(x+\omega^j)(x+\omega^{-j})\big],
\]
where the first $2$ factors are self-reciprocal and the remaining $6$ factors are ordered in reciprocal pairs. We have $s=2$ and $r=3$ such that $\G_i=\HH'_j=\HH''_j=\F_9$, for all $i\in\{1,2\}$ and $j\in\{1,2,3\}$. Consider the following constituents of length $2$ corresponding to the ordered factors:
\begin{align}\label{qutrit cons}
&C_1 : \begin{pmatrix}  1 & \omega^2 \end{pmatrix}, \, C_2  : 0_2, \, C_1' : I_2, \, C_1'' : 0_2, \notag \\
&C_2' :  0_2, \, C_2'' : \begin{pmatrix}  1 &  \omega^7 \end{pmatrix}, \, C_3' : \begin{pmatrix}  1 &  \omega^6 \end{pmatrix}, \, C_3'' : 0_2.
\end{align}

The QC code $C$ of index $2$ over $\F_9$ with those constituents has dimension $5$. Let $R=\F_9[x]/\langle x^8-1\rangle$. Then $C$ is generated by $\vec{f}_1(x)=(f_{1,0}(x), f_{1,1}(x)), \vec{f}_2(x)=(f_{2,0}(x), f_{2,1}(x))$ as an $R$-submodue in $R^2$, where \vspace{5pt}
\begin{align*}
f_{1,0}(x) = &~\omega^3 x^7 + \omega x^5 + \omega^6 x^3 + \omega^2 x - 1, \\
f_{1,1}(x) = &~\omega^2 x^7 - x^6 - x^4 + \omega^3 x^3 + \omega^7 x + \omega^3 \\
f_{2,0}(x)=&~0, \\
f_{2,1}(x) = &~\omega x^7 + \omega^6 x^6 + \omega^3 x^5 + x^4 + \omega^5 x^3 + \omega^2 x^2 +\\
&~\omega^7 x - 1.
\end{align*}
One can obtain the constituents listed in (\ref{qutrit cons}) by evaluating $\vec{f}_1(x)$ and $\vec{f}_2(x)$ at $-1, 1, \omega^5, \omega^3, \omega^6, \omega^2, \omega^7, \omega$, respectively.  We again have $e=1$, implying $\dim\left(C \cap C^{\bot_{\rm H}}\right)=4$. The Hermitian dual $C^{\bot_{\rm H}}$ is a $[16,11,5]_9$ code, which attains the best-known distance for a nonary code of this length and dimension. The minimum distance of $C + C^{\bot_{\rm H}}$ is $4$. Thus, by Theorem \ref{thm:X}, we obtain a $[\![17,7,5]\!]_2$ stabilizer code.

Proposition \ref{propagation} gives us two codes, with respective parameters $[\![17,6,5]\!]_3$ and $[\![18,7,5]\!]_3$. They are \emph{strictly better} than the $[\![17,6,4]\!]_3$ and $[\![18,7,4]\!]_3$ codes that held the previous record. The other derived code, with parameter $[\![16,7,4]\!]_3$, merely matches that of the current record holder.
\end{example}

\begin{table*}[t!] 
\caption{Generators of the Quasi-Cyclic codes used to construct the qutrit stabilizers in Table \ref{tab:qutrit}}
\renewcommand{\arraystretch}{1.3}
\setlength{\tabcolsep}{4pt}
\centering\resizebox{\textwidth}{!}{
\begin{tabular}{c l | c l}
\toprule
No & Generators & No & Generators \\
\midrule
1 &  $(1, \omega^2 \omega^4)$ &
			
26 & $(\omega^2 1 \omega^6 1 0 1 \omega^6 1 \omega^2 0,
\omega^3 \omega \omega^2 \omega^5 \omega \omega^5 \omega^2 \omega \omega^3 \omega^7),$\\
\cmidrule{1-2}
			
2 & $(1, \omega^7 \omega^3 \omega^7 \omega^7)$ &
			
& $(0, \omega^3 \omega^5 \omega \omega^7 1 \omega^7 \omega \omega^5 \omega^3 \omega^4)$ \\  
			
\midrule
			
3 &  $(\omega^6 1 \omega^2 0, \omega^5 \omega^3 \omega^2 1)$ &
			
27 & $(\omega^5 0 \omega^7 0 1 0 \omega^7 0 \omega^5 \omega^4,
\omega^3 \omega^4 \omega \omega^4 \omega \omega^3 \omega^5 0 \omega^7 \omega^6)$\\ 
\midrule
			
4 & $(\omega^3 \omega \omega \omega^3 1, \omega^6 \omega^5 \omega \omega^2 0)$ &
			
28 & $(\omega^3 \omega^6 \omega \omega^2 1 \omega^2 \omega \omega^6 \omega^3 1,
\omega^7 \omega^2 1 0 \omega^6 0 1 \omega^2 \omega^7 \omega^5)$\\ 
\midrule
			
5 & $(\omega^4 1 \omega^4 \omega^4 0 1 0 1,
\omega^4 \omega^5 \omega^7 \omega^2 \omega^3 0 \omega^3 \omega^7)$ &

29 &$(\omega^4 0 \omega^4 \omega^4 \omega^4 0 0 0 \omega^4 0 0, \omega^2 \omega^6 \omega^5 0 \omega^2 \omega \omega^5 \omega^6 \omega^3 \omega^5 \omega^4)$\\
\midrule
	
6 &  $(\omega^3 1 \omega 1 \omega^5 1 \omega^7 0,
\omega^2 1 \omega^5 \omega^5 \omega^4 \omega^5 0 \omega^6), (0, \omega^5 \omega^6 \omega^7 1 \omega \omega^2 \omega^3 \omega^4)$ &
			
30 & $(\omega^3 \omega^7 \omega 1 \omega^5 \omega^5 \omega^7 1 \omega^3 \omega^5 \omega 1
  \omega^5 \omega^7 \omega^7 0,$\\
\cmidrule{1-2}
			
7 & $(\omega^6 1 \omega^2 1 0 1 \omega^2 1 \omega^6 0,
\omega^3 0 \omega \omega^2 1 \omega^2 \omega 0 \omega^3 \omega^3),$ &
			
 & $\omega^7 \omega^3 \omega^7 \omega^7 \omega^3 \omega^3 \omega^7 \omega 0 \omega^5 \omega \omega^6 \omega^6 1 0 \omega^7)$\\
\cmidrule{3-4}
			
& $(0, \omega \omega^7 \omega^3 \omega^5 1 \omega^5 \omega^3 \omega^7 \omega \omega^4)$ &
		
31 & $(\omega^3 0 \omega \omega^7 \omega^7 0 \omega^5 1 \omega^3 \omega^7 \omega \omega^5
 \omega^7 \omega^5 \omega^5 \omega^4,$\\
\cmidrule{1-2}
			
8 & $(\omega^3 0 \omega 0 \omega^4 0 \omega 0 \omega^3 \omega^4, 
  \omega^7 0 0 \omega^3 \omega^5 \omega^4 \omega \omega^5 1 \omega), (0, \omega \omega^7  \omega^3 \omega^5 1\omega^5 \omega^3 \omega^7 \omega \omega^4 )$  & 

 & $\omega^3 \omega^7 \omega^3 \omega^5 \omega^7 \omega^4 \omega^2 \omega^7
 \omega^4 \omega^6 \omega^2 \omega^6 \omega \omega^2 \omega^4 \omega^5)$ \\
\midrule

9 & $(\omega^2 1 \omega^6 1 0 1 \omega^6 1 \omega^2 0, \omega^3 \omega^3\omega^7 \omega^3  \omega^3 \omega^7 \omega^5 \omega^2 \omega^5 \omega^7)$&
			
32 & $(1,\omega^7 \omega^5, \omega^4 \omega^5)$\\
\midrule
			
10 & $(\omega^7 0 \omega^5  0 1 0 \omega^5 0 \omega^7 \omega^4, \omega^4 0 \omega \omega^3  \omega^4 \omega^3 \omega^2 \omega^2 1 0)$  &

33 & $(\omega^4 \omega^4 \omega^4 0, \omega^3 \omega^3 \omega^7 \omega^7, \omega^7 \omega^4 \omega^5 \omega^4), (0, \omega^4 1 \omega^4 1, \omega^6 \omega^2 \omega^6 \omega^2)$\\ 
\midrule
			
11 & $(\omega^7 \omega^6 \omega^5 \omega^2 \omega^4 \omega^2 \omega^5 \omega^6 \omega^7 1, \omega^5 \omega^4 \omega^7 \omega 0 \omega \omega^7 \omega^4 \omega^5 \omega^5)$ &
			
34 & $(\omega^6 1 \omega^2 0, \omega^6 \omega^5 0 \omega^3, \omega^2 1 0 0)$\\
\midrule
			
12 & $(\omega^2 \omega \omega^6 \omega^3 \omega^4 \omega^3 \omega^6 \omega \omega^2 0, \omega^6 1 \omega^2 \omega  \omega^5 \omega^4 \omega^5 \omega \omega^2 1)$ &
			
35 & $(\omega 0 \omega^3 0 \omega^4 0 \omega^3 0 \omega \omega^4, \omega^2 \omega^2 \omega \omega^5 \omega \omega^2 \omega^2\omega \omega^5 \omega, \omega^4 \omega^7 \omega^7 1 \omega 0 \omega^5 \omega \omega^4 \omega^7),$\\
\cmidrule{1-2}
			
13 & $(1 0 0 0 1 1 1 0 1 1, 1 \omega \omega^3  \omega^4 \omega^7 \omega^6 \omega^2 0 0 \omega^5 0)$ &
			
 & $(0, \omega^5 \omega^3 \omega^7 \omega 0 \omega \omega^7 \omega^3 \omega^5 0, 
 \omega^3 \omega^4 1 \omega^7 \omega^5 \omega^7 1 \omega^4 \omega^3 \omega)$\\ 
\midrule
			
14 & $(\omega \omega^2 \omega^3 \omega^2 \omega^3 \omega^6 0 \omega^2 \omega \omega^6 \omega \omega^6 \omega^3 1, \omega^4 \omega^2 \omega^5 \omega^2 \omega^5 \omega^4 \omega^3 \omega^2 \omega^4 \omega^4 \omega^4 \omega^4 \omega^5 0)$ &
			
36 & $(\omega^4 \omega^6 \omega^4  \omega^2 \omega^4 \omega^2 \omega^4  \omega^6 \omega^4 \omega^4, 1 \omega \omega^4 \omega^6 \omega^3 \omega^7\omega \omega^4 \omega^2 \omega^2, $\\
\cmidrule{1-2}
			
15 & $(\omega^2 1 \omega^6 0, \omega^4 \omega^7 0 \omega^5, \omega^5 \omega^5 \omega^6 \omega^7)$ &
			
& $1 \omega^6 \omega^7 0\omega^2 \omega^3 \omega^3 \omega^3 \omega^5)$\\
\midrule 
			
16 & $(\omega^5 \omega^7 \omega^7 \omega^5 0, 1 \omega^7 \omega^3 \omega^4 0, \omega^4 1 \omega^3 \omega^3 1)$ &
			
37 & $(1, 0, \omega \omega^4, \omega^5, \omega^7 \omega), (0, 1, \omega^6 \omega^5, 1 \omega, \omega^3 \omega^6)$\\
\midrule
			
17 & $(\omega^6 \omega \omega^2 \omega^3 1 \omega^3 \omega^2 \omega \omega^6 0, \omega^2 \omega^7 \omega^5 0 \omega^7 0 \omega^5 \omega^7 \omega^2 1, \omega \omega \omega^5 \omega^5 \omega^4 \omega^7 \omega^2 \omega^6 \omega^3 1)$ &
			
38 & $(\omega^5\omega^7\omega^7\omega^5 0, \omega^2\omega^2\omega^2\omega^2\omega^2, \omega^2\omega^6 \omega^3 1, \omega^2 0\omega^3 \omega\omega^7, \omega \omega^3 0 \omega^2 \omega^7),$ \\
\cmidrule{1-2}
			
18 & $(1, \omega^6 \omega^7, \omega^3 \omega^6, \omega^4 \omega^6)$ &
			
 & $(0, \omega^3\omega\omega\omega^3 1, \omega^3 1\omega^3 \omega \omega, \omega^3 \omega^3 \omega^5\omega^2\omega^5, \omega \omega^3 1 \omega^3 \omega)$\\
\midrule
			
19 & $(1,0,\omega^2\omega^7,\omega^2\omega^3,\omega^4\omega^6), (0, 1, 11, \omega^2\omega^6, \omega^6 0)$ &
			
39 & $(\omega^5\omega^7\omega^7\omega^5 0, 1\omega^7\omega^3\omega^4 0), (0, \omega^4\omega^4\omega^4\omega^4\omega^4)$\\
\midrule
			
20 & $(\omega^5\omega^7\omega^7\omega^5 0, \omega^7\omega^7\omega^7\omega^7\omega^7, \omega^3\omega^7\omega^7\omega^3\omega, \omega\omega^2 1 \omega^4\omega^3, 1\omega^3\omega^3 1\omega^4),$ &
			
40 & $(\omega^6 0 \omega^2 0 \omega^3 0 \omega\omega^4, \omega^6 \omega^7 0 1 \omega^4 \omega^2 \omega^7 \omega^7), (0, \omega^3 \omega^3 \omega 0 \omega^2 \omega \omega^6 1)$\\
\cmidrule{3-4}
			
 & $(0, \omega^3\omega\omega\omega^3 1, \omega^2\omega^2\omega^4\omega\omega^4, 1 \omega^4\omega^5 \omega 0, \omega^6\omega^3 \omega^6\omega^4\omega^4)$ & 
			
41 & $(1 0 1 0 \omega^5 0 \omega^7 \omega^4, \omega^2 \omega^7 \omega^5 \omega \omega 1 \omega^6 \omega), (0, 1 \omega^4 1 \omega^4 1 \omega^4 1 \omega^4)$ \\
\midrule
			
21 & $(\omega^7\omega^5\omega^5\omega^7 0, 1\omega^5\omega^2\omega^5 1,  \omega^4 0 1\omega^5 \omega, \omega\omega^4 \omega 0, 1\omega^7\omega^4\omega^4\omega^7)$ &
			
42 & $(\omega \omega \omega^3 0 \omega^6  \omega^3 \omega^2 1, \omega^5 1 \omega^3 \omega^6 \omega \omega^4 \omega^7 \omega^2), (0, 1 \omega^4 1 \omega^4 1 \omega^4 1 \omega^4)$\\
\midrule
			
22 & $(1, 1\omega^7)$ & 
			
43 & $(\omega^2 1 \omega^6 1 0 1 \omega^6 1 \omega^2 0,  \omega \omega^2 1 1 \omega^6 
 \omega^5 \omega^2 \omega^3 \omega^4 \omega^5), (0, \omega^4 1 \omega^4 1 \omega^4 1 \omega^4 1 \omega^4 1)$\\
\midrule
			
23 & $(\omega^6 1 \omega^2 0, \omega^2 \omega^2 \omega^5 1)$ &
			
44 & $(\omega^3 \omega^6 \omega \omega^2 1 \omega^2 \omega \omega^6 \omega^3 1, 
  \omega^3 \omega \omega^4 \omega^5 \omega^5 \omega^4 \omega \omega^3 0 0), (0, \omega^4 1 \omega^4 1 \omega^4 1 \omega^4 1 \omega^4 1)$\\ 
\midrule

24 & $(\omega^7 \omega^5 \omega^5 \omega^7 0, \omega^7 \omega^7 \omega^4 \omega \omega^4)$&
			
45 & $(\omega 0 \omega^3 \omega^4, \omega^3\omega^5\omega^7\omega, \omega^4\omega^3\omega^3\omega^6, \omega^2 1 \omega^7\omega^7, \omega^6\omega^4\omega^3\omega^3),$ \\
\cmidrule{1-2}

25 & $(\omega^6 0 \omega^2 1 \omega^2 0 \omega^6 0,
\omega^2 0 \omega^3 \omega^2 \omega \omega \omega \omega^4), (0, \omega^7 \omega^2 \omega^5 1 \omega^3 \omega^6 \omega \omega^4)$ &

& $(0, \omega^4 1\omega^4 1, \omega\omega^5\omega\omega^5, 1 \omega^4 1 \omega^4, \omega^7\omega^3\omega^7\omega^3)$\\
\bottomrule
\end{tabular}}
\label{tab:cons}
\end{table*}

Table \ref{tab:qutrit} lists the qutrit codes that we have found through random search, performed using {\tt Magma}~\cite{Magma}, over the constituents that satisfy the requirements described in Section \ref{hull section}. We include some important parameters needed to construct the codes explicitly. Table \ref{tab:cons} contains the generating polynomials corresponding to each row in Table \ref{tab:qutrit}, where each polynomial $f(x)=a_{\deg(f)}x^{\deg(f)}+\cdots+a_1 x + a_0$ is represented by an array of its coefficients $a_{\deg(f)}\cdots a_1 a_0$. All of the listed codes are distinct from those in~\cite{Edel} and in~\cite[Sec.~5]{Galindo2018}.

%\bibliographystyle{IEEEtran}
%\bibliography{algcodes} 


\end{document}